# A Clustering Strategy for Enhanced FL-Based Intrusion Detection in IoT Networks

Jacopo Talpini,[1] Fabio Sartori,[1] and Marco Savi[1]

[1]*Department of Informatics, Systems and Communication, University of Milano-Bicocca, Milano, Italy*[*]

(Dated: February 2023)

The Internet of Things (IoT) is growing rapidly and so the need of ensuring protection against cybersecurity attacks to IoT devices. In this scenario, Intrusion Detection Systems (IDSs) play a crucial role and data-driven IDSs based on machine learning (ML) have recently attracted more and more interest by the research community. While conventional ML-based IDSs are based on a centralized architecture where IoT devices share their data with a central server for model training, we propose a novel approach that is based on federated learning (FL). However, conventional FL is ineffective in the considered scenario, due to the high statistical heterogeneity of data collected by IoT devices. To overcome this limitation, we propose a three-tier FL-based architecture where IoT devices are clustered together based on their statistical properties. Clustering decisions are taken by means of a novel entropy-based strategy, which helps improve model training performance. We tested our solution on the CIC-ToN-IoT dataset: our clustering strategy increases intrusion detection performance with respect to a conventional FL approach up to +17% in terms of F1-score, along with a significant reduction of the number of training rounds.

## I. INTRODUCTION

The Internet of Things (IoT) is an emerging paradigm that enables the interconnection of heterogeneous devices and computing capabilities in the Internet. However, the continuous development of IoT systems including a large number of devices leads to an increased risk of cyber-attacks, therefore security and privacy are widely considered critical issues in such a context [7].

One of the main security measures in modern networks are Intrusion Detection Systems (IDSs), whose aim is to identify attacks, unauthorized intrusions, and malicious activity in networks [30]. The traditional approach for detecting intrusion relies on knowledge-based systems [7] but as long as networks rise in complexity they become more prone to error [26, 30]. As a consequence, data-driven approaches based on machine learning (ML) have been widely considered in the recent years for the detection of attacks, also in IoT scenarios [1].

However, a limitation of current ML-based approaches is that model training is based on data and computational power elaborated and owned by a centralized node (e.g. a server). Centralized ML approaches are thus generally associated with different challenges including the need of high computational power and long training time, as well as with the rise of security and privacy concerning users' data [18].

[*] {name.surname}@unimib.it

In order to address these issues, federated learning (FL) was originally proposed in [17] and has recently emerged as an effective model training paradigm to address the issues recalled above. It embodies the principles of *focused collection* and *data minimization*, and can especially mitigate many of the systemic privacy risks and costs resulting from traditional, centralized machine learning, including high communication efficiency and low-latency data processing [11].

In an IoT environment, devices typically collect data samples that are not independent and identically distributed (iid) [4]. This scenario, known as *statistical heterogeneity* of the collected clients' (i.e., IoT devices') datasets, poses a challenge for a federated learning setting [13] and is particularly detrimental when developing and deploying an FL-based IDS, since some clients may have traffic associated with several kinds of attacks (e.g. DoS, port scanning, etc.), while other could only have traffic related to their intended operation [4].

To get around the problem of imbalance between IoT devices' data we propose a three-tiered FL-based system where a set of *data aggregators* act as FL clients, each on behalf of a group (or *cluster*) of IoT devices. IoT devices are clustered by means of a novel strategy employing a *similarity score* to measure the statistical similarity of high-level attacks' data collected by any IoT device, and assign *similar* IoT devices (in terms of suffered attacks) to different clusters. In this way attacks' labeled samples, as employed by the data aggregators for model training, are re-balanced with a consequent enhancement of the FL-based model training procedure.

We tested the validity of our approach on CIC-ToN-IoT dataset [23], one of the latest and more adopted dataset related to attacks towards an IoT infrastructure [2]. We compared it to a classical FL approach, where IoT devices act as FL clients (no matter how training data is distributed), and a centralized approach, where the model for intrusion detection is trained with data collected from IoT devices in a centralized location. Our results show that our proposal is effective and can enhance intrusion detection performance with respect to a classical FL approach, leading to results close to those obtained in a centralized setting.

To summarize, our main contributions are:

- The definition of a novel FL-based three-tier system for model training and clustering, for enhanced intrusion detection in IoT devices.

- The definition of a similarity score for measuring the overall imbalance between IoT devices' data.

- The definition of a clustering strategy driven by the optimization of our score.

- An evaluation of how different cluster configurations impact the overall attacks' classification performance.

The structure of the paper is organized as follows. Section II introduces the related work, while Section III describes the system architecture and Section IV introduces the proposed clustering strategy. Finally, Section V describes the experiments and related results, while Section VI concludes the paper.

## II. RELATED WORK

Intrusion Detection Systems (IDSs) are traditionally considered as a second line of defence, with the aim of monitoring the network traffic and detecting malicious activities that have eluded the security perimeter [19]. IDSs are generally divided into signature-based or anomaly-based [29]. The first category, also known as misuse IDS, is based on pattern recognition, with the goal of comparing signatures of well-known attacks to current network traffic patterns. On the other hand, anomaly-based methods rely on a model for the normal network traffic so that any pattern that deviates from the usual one is considered an intrusion.

In this paper, we focus on signature-based intrusion detection for IoT, and in the following subsection we report on relevant work related to data-driven systems, based on machine learning, in this context. Later we also recall relevant recent strategies that have been proposed for improving performance in a FL setting, including clustering of FL clients.

### A. ML-Based Intrusion Detection Systems

In recent years, data-driven approaches for developing IDSs have been explored [15, 26, 29] considering different methods such as random forests, support vector machines, neural networks or clustering techniques. In particular, machine learning and deep learning are emerging as promising data-driven methods with the capability to learn and extract harmful patterns from network traffic, which can be beneficial for detecting security threats occurring in networked systems in general [26], and on IoT networks in particular [5, 22]. However, the vast majority of the proposed IDS approaches adopted in the IoT domain that can be found in the literature relies on centralized approaches where IoT devices send their local dataset to cloud datacenters (or centralized servers) to leverage on their computing capabilities for model training [4].

As a consequence, the specification of a ML-based IDS whose model is trained following a federated learning approach seems a very promising alternative solution in such a domain, and it is possible to find a few examples in the literature exploring its feasibility [3, 4, 20, 21]. For instance, in [21] the authors presented a framework that uses a federated learning approach to detect malware, based on both supervised and unsupervised models. More precisely, the authors perform a binary classification with a multi-layer perceptron and an autoencoder on balanced datasets, which present the same class proportions for every client. Another relevant contribution is represented by the paper [4], in which the authors investigated a FL approach on a realistic IoT dataset (i.e., CiC-TON-IoT) [2]. In particular, the authors evaluated how the unbalance between clients' data class distribution affects the performance of the federated learning workflow. They also suggest the usage of Shannon entropy to measure the unbalance within each client data. With respect to [4], in this paper we define a more general score for assessing the unbalance of the data distribution and we leverage it to design an architecture where IoT devices' data is balanced without any need of sharing samples between themselves, assumption that is instead made in that paper.

### B. Approaches for Enhancing Federated Learning

Despite FL has been recently introduced, it has received a lot of interest from the research community [13]. The standard FL workflow is based on the *FedAvg* algorithm [17], which essentially relies on a weighted average of the models received from each client after local training. A crucial aspect in a FL setting is the statistical heterogeneity of the client datasets, which can be harmful to training performance. In literature, many works can be found that aim at analyzing this issue and propose possible solutions [9, 13, 24].

A possible approach to solve the issue consists of learning a personalized model for each FL client. For instance, the MOCHA algorithm proposed in [28] considers a multi-task learning setting and defines a deterministic optimization problem where the correlation matrix of the client is exploited as a regularization term. Another possible approach is to consider FL with heterogeneous data as a meta-learning problem. In this scenario, the goal is to obtain a single global model, and then let each client fine-tune it using its local data [10].

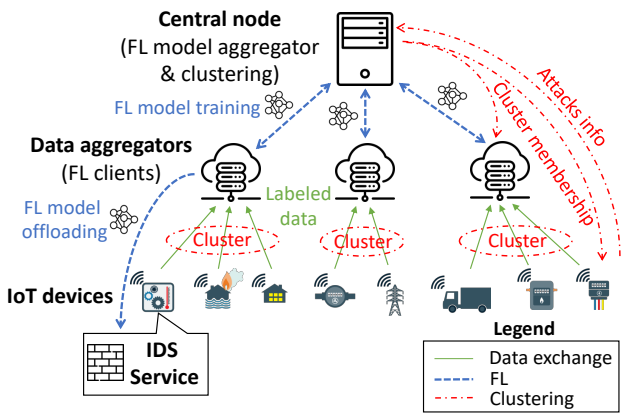

FIG. 1. FL-based intrusion detection with IoT devices' clustering: system architecture.

Lastly, it is possible to tackle this issue by considering a *clustered* FL framework [6, 24]. For instance [24] proposed a centralized clustering algorithm based on the geometric properties of the FL loss surface, so that each client may deploy a more specialized model. The proposed procedure is applied after FL training, once it has converged to a stationary point, which may lead to high computational costs [6]. Another example is given by [25], where K-Means clustering is leveraged for grouping FL clients according to some common features and then collaboratively training an FL model per cluster, rather than deploying a single global model. The paper shows that this leads to an increase in performance in the inference phase. However, in our context, it may be unwise to let each client (or a cluster of them) deploy a specialized model since it would be desirable for each of them to be able of recognizing also new attacks, seen by other IoT devices. In this paper, we try to fill this gap by proposing a three-tier clustering architecture where a single global model is trained, and IoT devices are clustered together based on a novel similarity score.

However, unless its very simple and fast applicability, FedAvg may not be an optimal choice in some cases, e.g. for clients with highly unbalanced and non iid data distributions, due to convergence issues [24]. Moreover, communication is a critical bottleneck in federated networks and reducing the total number of communication rounds is a key aspect to be considered [13]. In the next section, we describe our proposed architecture to deal with these shortcomings.

## III. SYSTEM ARCHITECTURE

Fig. 1 shows our proposed architecture for FL-based intrusion detection. It consists of a three-tier system including *IoT devices*, *data aggregators* and a *central node*. These components are responsible for any of the main tasks performed by the system: (i) *FL-based model training* (blue arrows in the figure), (ii) *labeled data exchange* (green arrows), and (iii) *IoT devices clustering* (red arrows).

With respect to existing FL-based architectures for intrusion detection [4], the main difference is the presence of the data aggregators. These nodes act as FL clients and, during the FL-based model training procedure that involves themselves and the central node (i.e., FL model aggregator) executing FedAvg, they use labeled traffic data gathered from a subset of clustered IoT devices, which is opportunely aggregated. Traffic data from any IoT device consists of a set of historical attack/benign traffic samples that should have been collected and labeled by the owner/administrator of such an IoT device.

Unlike [4], the training data can be balanced among the nodes in charge of model training (i.e., the data aggregators) without the need of exchanging part of these data among IoT devices, which is cumbersome in most IoT real-world scenarios and could lead to privacy issues if IoT devices are owned by different parties. In addition, IoT devices are relieved from any training task, which is often very demanding from a computational point of view given their hardware equipment. Then, each data aggregator offloads the trained *FL model* to the IoT devices it is responsible for, so that it can be used for intrusion detection (i.e., inference) on live traffic by the IDS service implemented in any IoT device.

A fundamental aspect of our proposed system is thus how IoT devices are clustered and consequently associated to any data aggregator. This is key to ensure that local FL models are trained on likely balanced datasets so that overall intrusion detection performance is improved. Details on our clustering procedure are provided in Section IV: from an architectural point of view, this procedure is carried out by the central node with some high-level attack information provided by the IoT devices. Such an information, as we will better describe in Section IV, includes the statistical distribution of the attacks experienced by the IoT devices in the past.

Once the clusters' composition is computed by the central node, the cluster membership of each IoT device is communicated to its associated data aggregator and to the IoT device itself, so that (i) traffic labeled data can be sent from the IoT device to the designated data aggregator for FL-based model training and (ii) the trained FL model can be offloaded from the designated data aggregator to the IoT device for local intrusion detection.

### A. Example scenario

Our system could be adopted in different scenarios. As a matter of example, let's consider an Internet Service Provider (ISP) that wants to offer an IDS service to its customers, each one owning a bunch of IoT devices at the far edge (e.g. for smart agriculture [27]). In this specific case, the data aggregators' logic can be implemented by the edge computing nodes [12] owned by the ISP and disseminated at the edge of its infrastructure. In addition, the central node can be implemented in the cloud datacenter managed by the same ISP. In this specific case, the FL model used for inference by the IoT devices can be trained in the cloud-to-things continuum while ensuring its best possible quality, regardless of how balanced or imbalanced the labeled attack traffic data from any single IoT device is.

Moreover, performing federated learning between edge computing nodes and the central cloud can help reduce the amount of data that needs to be moved around in the ISP network if compared to a *centralized* solution, where model training is fully executed in the cloud [25]. In fact, only model weights need to be transferred between edge computing nodes and the cloud, while labeled data related to attacks needs only to be moved from IoT devices to designated edge computing nodes, and is thus kept as much as possible local. Note also that sharing the labeled attacks' data with edge computing nodes should not be considered a potential source of privacy leakage, as this data is kept within the ISP network domain offering the IDS service.

## IV. CLUSTERING IOT DEVICES: A NOVEL STRATEGY

To define our strategy, we took inspiration from some findings reported in [4]. In that paper, the authors showed that FL works well in a *balanced scenario*, when all the clients have the same number of samples for each class.

Here we propose a *clustering strategy* with the aim of automatically choosing an optimal configuration of the clients partition in each cluster, by analyzing their labels distributions. The aim is to find a meaningful score so that, for a given number of clusters, its minimization results in an effective clusters' configuration, which can then be deployed in the three-tier architecture described in Section III.

For each client dataset of size $N$, given the total number of classes $K$ (i.e., benign and/or attack traffic) and the number of per-class samples $n_i$, it is possible to compute the normalized Shannon entropy of the labels probability distribution $= \{p_1; ...; p_K\}$ as [16]:

$$\mathrm{H}(p) = \frac{-\sum_{i=1}^{K} p_i \ln(p_i)}{\ln(K)} \quad (1)$$

where $p_i$ is the probability of a sample belonging to the $i$-th class to occur, which can be expressed as $p_i = n_i/N$. The Shannon entropy is a measure of the uncertainty associated with a given probability distribution and it is maximized when the distribution is flat [16] (i.e., when all classes have the same number of samples).

This is similar to what was done in [4]. However, groups with the same entropy may have different labels distributions, so it is necessary to take into account also their similarity. For instance, two IoT devices where all the traffic of any of them belongs to one class, which is mutually exclusive with respect to the class of the other device, have the same entropy (equal to zero) but provide substantially different information about the experienced attacks.

Some of the most common choices for measuring the similarity between two probability distributions are: (i) the Kullback-Leibler divergence; its symmetrized extensions, such as (ii) the Jensen and Shannon distance [14], and (iii) the Hellinger distance [8]. Here we decided to exploit the latter because it satisfies some desirable properties: it is symmetric, it satisfies the triangle inequality and it lies in the range $[0, 1]$.

The Hellinger distance between two discrete probability distributions $p = \{p_1; ...; p_K\}$ and $q = \{q_1; ...; q_K\}$ is defined as:

$$d(p,q) = \frac{1}{\sqrt{2}} \sqrt{\sum_{i=1}^{K} (\sqrt{p_i} - \sqrt{q_i})^2}. \quad (2)$$

It may be desirable to have clusters where each group has a high entropy and that all groups pairs have a low distance between their labels distribution. In this way, it is ensured that training data among clusters is balanced as much as possible. Starting from these considerations we define a *similarity score* for ranking the overall balance provided by different groups/clusters[1] configurations. Denoting a clustering configuration of $N$ groups as $R = \{p_1; ...; p_N\}$, we define the clustering score as:

$$S(R) = \frac{1}{N} \sum_{i=1}^{N} \frac{1}{\mathrm{H}(p_i)} + \frac{1}{2\binom{N}{2}} \sum_{\substack{i,j=1 \\ i \neq j}}^{N} d(p_i, p_j). \quad (3)$$

The first term takes into account the entropy of each group and heavily punishes configurations in which even just one group has low entropy, as such a condition is strongly in contrast with a balanced scenario. The second term describes the dissimilarity of the labels distribution between groups. The sum runs over the possible groups' pairs, since the distance is symmetric we included a factor $1/2$ which accounts for adding each pair of groups twice while the factor $\binom{N}{2}$ represents the number of pairs combinations.

Therefore, for a fixed number of clusters $N$, the clustering configuration that minimizes $S$, as defined in Eq. (3), should be chosen. The number of groups can be considered as a parameter of the proposed clustered federated learning workflow. It has to be properly tuned by taking into account also the physical constraints of the considered network, like the number of aggregators, in order to achieve the best trade-off between possible performance gains obtained by clustering and the amount of devices' traffic data that needs to be aggregated. In Eq. (3) all the possible pairs of clusters have been considered, which may lead to a computational issue for a large number of clusters. However, it is expected that the best performance is achieved with a few number of clusters, which makes the computation is feasible: this will be evident from the evaluations of the next section.

---

[1] In this paper, the terms *groups* and *clusters* are used interchangeably.

## V. EXPERIMENTS AND RESULTS

### A. Description of the Dataset

For exploring the feasibility of the adoption of a federated learning approach for the detection of intrusions in IoT networks, the choice of a realistic dataset plays a crucial role. In [4] it is possible to find an extensive review of different existing datasets. As done in that paper and given its properties, here we exploit the CIC-ToN-IoT dataset [23], which is based on the ToN_IoT set [2]. The dataset includes real collected data and is based on 83 features describing the network traffic between different source/destination IP addresses. It is organized *per flow*: each row represents a flow and is annotated as belonging to one over a total of 10 classes, including benign traffic and nine different attacks.

From the dataset we selected the 16 destination IP addresses, each associated with a different IoT device, with more samples. For each of them, we considered also the flows related to transmitted traffic (i.e., when they act as sources) so that each IoT device's dataset consists of received and sent traffic, mimicking a realistic scenario. After this operation, the overall length of our dataset (not partitioned) was 9849282 samples. We added a binary feature that takes into account the direction of the traffic with respect to an IoT device and we dropped some device-specific features such as the IP address and the flow timestamp. The resulting dataset is unbalanced and class distributions are highly dissimilar across the clients.

Lastly, each IoT device's dataset is divided into train, validation and test set, 60%, 20%, 20% respectively, and standardized using the training data, so that each feature distribution presents a zero mean and unit variance. The validation set is exploited for tuning the hyper-parameters of the classifier and of the federated learning setting (batch size, learning rate, number of local epochs and number of rounds), and the architecture of the classifier. In the next subsections we recall settings related to the chosen classifier, to the baseline approaches and to our proposed approach.

### B. Adopted Classifier and Baseline Approaches

As we are interested in characterizing the impact of clustering clients in different ways, we selected a simple classifier, represented by a multi-layer perceptron (MLP) consisting of an input layer, a single hidden layer with 64 neurons, and a final outer layer. The hidden layer makes use of a hyperbolic tangent activation function while the outer of Softmax. The optimization of each client is performed using the Adam optimizer with an initial learning rate of 0.001 and the cross-entropy as loss function; each MLP is trained for up to 15 epochs with batches of 512 samples. The federated learning training workflow varies depending on the considered approach, as described below. In every case we simulated the FL training procedure using TensorFlow Federated, version 0.19.

The first baseline approach is represented by the usual federated learning approach, where the 16 selected IoT devices are the FL clients, which train local models using locally-collected data. In this specific case, no data aggregators are included in the related architecture, which only consists of FL clients and of a centralized aggregator. In this scenario we trained the global model for 200 consecutive rounds using FedAvg as aggregation function. As no clustering is performed in this case, we will refer to this baseline approach as *no groups*.

The second baseline approach is a centralized scenario where each client transmits its own data to a central server, which thus has a global dataset at its disposal. Then, the classifier is trained simultaneously on the global dataset with the previously described parameters and hyper-parameters. We refer to this second baseline approach as *centralized*.

### C. Proposed Approach

We experimented with different numbers of configurations, with the goal of verifying that our proposed similarity score represents an effective metric for IoT devices' clustering. We considered three *cases*, associated with different $N$ values: (i) $N = 2$, i.e., two clusters with eight IoT devices each, (ii) $N = 4$, i.e., four clusters with four IoT devices each, and (iii) $N = 8$, i.e., eight clusters with two IoT devices each. Given $N$, the choice of how to partition the IoT devices and assign them to data aggregators follows the scheme reported in Algorithm 1. We randomly chose different partitions of clients obtaining different $K$ realizations of the clusters. For each of them, we computed the score defined in Eq. (5), and then we trained our model in a federated learning way for the best and the worst group realizations according to the score. This procedure is repeated for all the different $N$ values defined above.

**Algorithm 1** Clusters selection strategy
---
**Require:** Labels distribution of each of the $N$ randomly-generated clusters $\{p\}_{j=1}^{N}$
**Require:** $K$ random realizations $\{R \equiv [p_1,..,p_N]\}_{k=1}^{K}$ of the clusters
**Require:** Array for the $K$ clusters scores $S$
    **for** $k = 1,..,K$ **do**
    $S_k = S([p_1,..,p_N]_k)$      ▷ as in Eq.(5)
    **end for**
    Let $S^* = \text{sort}(S)$
    $S^*[0] \to$ Best groups realization ($R_{\text{best}}$)
    $S^*[K] \to$ Worst groups realization ($R_{\text{worst}}$)
    **return** $R_{\text{best}}, R_{\text{worst}}$

### D. Results

For each of the considered cases, we tested the trained classifier by means of FL on each client test set. We considered the *F1-score* as a metric for assessing the performance of the classification on a single client. All the experiments were performed on a Fujitsu workstation with 32 GB of RAM, equipped with an Intel Core i7 and a NVIDIA Quadro P1000 GPU.

In Table 1, for each considered configuration obtained by running Alg. 1 and for the baseline strategies, we report the arithmetic mean of F1-score computed over all the 16 IoT devices, along with the standard deviation (STD). We decided to exploit the arithmetic mean rather than the most commonly used weighted mean (over the number of clients' samples) since a few IoT devices are associated with a very large amount of benign traffic: considering a weighted mean would lead to an overestimated (and thus optimistic) performance.

TABLE I. Test performance for different training groups configurations.

| Configuration | Similarity score | F1 Mean | F1 STD |
|---|---|---|---|
| Centralized | - | 0.891 | 0.110 |
| $N = 2$ (2 groups) | 2.11 (Lowest) | 0.859 | 0.123 |
| $N = 2$ (2 groups) | 2.51 (Highest) | 0.828 | 0.141 |
| $N = 4$ (4 groups) | 2.16 (Lowest) | 0.856 | 0.132 |
| $N = 4$ (4 groups) | 2.68 (Highest) | 0.797 | 0.156 |
| $N = 8$ (8 groups) | 2.53 (Lowest) | 0.814 | 0.146 |
| $N = 8$ (8 groups) | $+\infty$ (Highest) | 0.769 | 0.167 |
| No groups | $+\infty$ | 0.730 | 0.207 |

Moreover, in Fig. 2 we present a graphical representation of the distribution of IoT devices F1-scores through violin plots. Violin plots feature a kernel density estimation of the underlying distribution: the shape of the violin plots represents a visual aid on the behavior of the classifier as a function of the groups' configuration along with the considered baseline models. Violins with denser tails represent better stability: in this case, the groups' configuration allows the trained classifier to produce fewer outliers with low F1-score during the test phase.

From Table 1 and Fig. 2 it is possible to notice that our strategy is better than the standard federated learning approach (*no group*) in terms of F1-score, and thus that the proposed score can be effectively used to assess groups unbalance. This is reflected in systematically higher average F1-scores and low STDs for configurations that lead to a lower similarity score, which are configurations with 4 and 2 clusters. In these settings performance are comparable, however the configuration with four groups should be preferred since it would increase the granularity offered by our system and hence it would reduce the expected computational overhead of aggregators with respect to the configuration with two clusters. The best F1-score is however obtained by *centralized*, as all classes samples are present during the training of the global model.

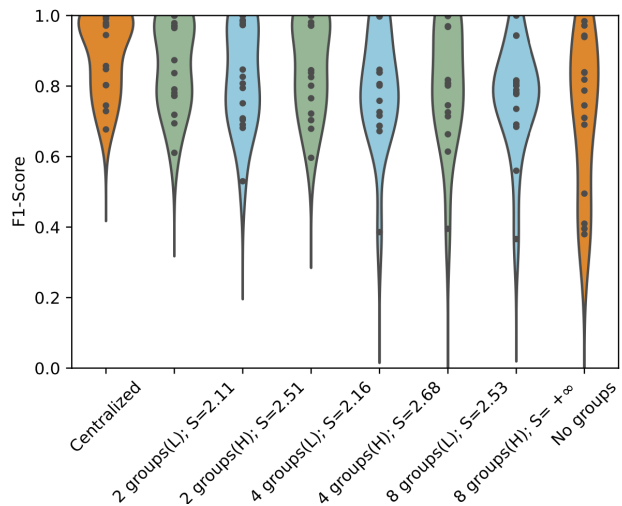

FIG. 2. Violin plots of test performance (L=Lowest, H=Highest). The y-axis reports the F1-score for each IoT device test set. The violin plots represent a pictorial behaviour of the test performance for a given cluster configuration. Violins with denser tails at higher F1-Score are considered better choices.

TABLE II. Test performance on new 35 clients, for different training groups configurations.

| Configuration | Similarity score | F1 Mean | F1 STD |
|---|---|---|---|
| Centralized | - | 0.867 | 0.151 |
| $N = 2$ (2 groups) | 2.11 (Lowest) | 0.849 | 0.153 |
| $N = 2$ (2 groups) | 2.51 (Highest) | 0.831 | 0.176 |
| $N = 4$ (4 groups) | 2.16 (Lowest) | 0.853 | 0.155 |
| $N = 4$ (4 groups) | 2.68 (Highest) | 0.819 | 0.177 |
| $N = 8$ (8 groups) | 2.53 (Lowest) | 0.827 | 0.162 |
| $N = 8$ (8 groups) | $+\infty$ (Highest) | 0.658 | 0.201 |
| No groups | $+\infty$ | 0.621 | 0.204 |

It should also be noted that, in our strategy, the training rounds can be significantly reduced with respect to *no group*. While in *no groups* 200 training rounds were performed, when clustering is adopted we could run the federated learning training for only 90 communication rounds, after which the overall training progress did not improve anymore.

Lastly, we tested our models on new 35 IoT devices' datasets, not included in the training phase and randomly chosen from the remaining devices in the CiC-ToN-IoT dataset. This new dataset has an overall length of 175726 samples. Table 2 shows the results for this test, which essentially confirm the performance trends inferred from Table 1. Moreover, it also shows that the proposed approach can train ML models that are able to generalize well to new IoT devices' data. This is a relevant aspect, especially for practical applications, since it may happen that some IoT devices are not able to share their data labels due to stringent privacy policies or because they have not collected enough historical samples yet.

## VI. CONCLUSION

In this paper, a federated-learning-based Intrusion Detection System for IoT networks, based on IoT devices' clustering, has been presented. In summary, the proposed approach relies on a new clustering similarity score, which is a meaningful quantity for grouping IoT devices and its minimization is beneficial in a FL training scenario both in terms of classification accuracy and number of rounds. More precisely our approach increases intrusion detection performance with respect to a conventional FL approach up to +17% in terms of F1-score with a halved number of training rounds. Moreover, the classifier trained with the proposed framework is able to generalize well to new unseen IoT devices, thus effectively sharing the knowledge of the attacks to new devices without sharing local data.

Here we have exploited a random search for choosing the clustering configuration, in future works we plan to develop also a custom algorithm for finding the near-optimal clustering score for a given collection of clients and number of clusters, as long as new tests on different datasets. Moreover, we plan to further investigate the effects of other aggregation algorithms that can overcome FedAvg limitations in unbalanced data scenarios, as well as specific techniques for dealing with unrepresented attacks classes to improve the overall classification performance.